\documentclass[aps,prl,twocolumn,showpacs,reprint,groupedaddress,superscriptaddress,longbibliography]{revtex4-1}

\usepackage[english]{babel}
\usepackage{float}
\usepackage[caption=false]{subfig}
\usepackage{amsfonts}
\usepackage{amssymb}
\usepackage{mathbrack}
\usepackage{siunitx}
\usepackage{mathtools}
\usepackage{bbold}
\usepackage{physics}
\usepackage{mwe}
\usepackage{amsthm,enumitem}
\usepackage{tikz}
\usepackage{ulem}
\usetikzlibrary{arrows,arrows.meta,shapes}
\usetikzlibrary{positioning}
\pdfoutput=1
\usepackage{dcolumn}
\usepackage{bm}
\usepackage[utf8]{inputenc}
\usepackage{graphics,graphicx}
\usepackage{parskip}
\usepackage{amsmath,amssymb}
\usepackage{multirow}
\usepackage{fancyhdr}
\usepackage{xcolor}
\usepackage{placeins}
\usepackage[title]{appendix}
\usepackage{wasysym}
\usepackage{url}
\usepackage{braket}
\usepackage{siunitx}
\usepackage{upgreek}
\usepackage[caption=false]{subfig}

\begin{document}

\title{Isospectrally Patterned Lattices}

\author{Peter Schmelcher}
\email{peter.schmelcher@uni-hamburg.de}
\affiliation{Zentrum f\"ur Optische Quantentechnologien, Fachbereich Physik, Universit\"at Hamburg, Luruper Chaussee 149, 22761 Hamburg, Germany}
\affiliation{The Hamburg Centre for Ultrafast Imaging, Universit\"at Hamburg, Luruper Chaussee 149, 22761 Hamburg, Germany}
\affiliation{ITAMP, Center for Astrophysics $|$ Harvard $\&$ Smithsonian Cambridge, Massachusetts 02138, USA }

\date{\today}

\begin{abstract}
We introduce and explore patterned lattices consisting of coupled isospectral cells that vary
across the lattice. The isospectrality of the cells is encapsulated in the phase that 
characterizes each cell and can be designed at will such that the lattice exhibits a certain
phase gradient. Focusing on the specific example of a constant phase gradient on
a given finite phase interval we show that the resulting band structure consists of three
distinct energy domains with two crossover edges marking the transition from single center
localized to delocalized states and vice versa. The characteristic localization length
emerges due to a competition of the involved phase gradient
on basis of a local rotation and the coupling between the cells 
which allows us to illuminate the underlying localization mechanism and its evolution.
The fraction of localized versus delocalized eigenstates can be tuned by changing the phase gradient between
the cells of the lattice. We outline the perspectives of investigation of this novel class of isospectrally
patterned lattices.
\end{abstract}

\maketitle

\paragraph{Introduction -} 
Symmetries are ubiquituous in our description of quantum matter and represent
a powerful means to analyze and classify its properties \cite{Hamermesh89}. They define a unique
starting-point for a subsequent deductive analytical or numerical study, a 
famous example being the theory of band structure which is based on 
Bloch's theorem due to crystalline translation invariance \cite{Ashcroft76,Singleton01}. The latter implies
completely delocalized Bloch states. The absence of any symmetries in the case of disorder leads in one 
and two spatial dimensions to a spectrum of localized states \cite{Lee85,Brandes03}. Quasicrystals
with their aperiodic long-range order fall into the substantial gap between these two limiting cases 
\cite{Macia09,Macia21,Shechtman84,Suck02,Janssen86,Berger93,Vieira05,Tanese14,Jagannathan21,Morfonios14}.
They show fractal energy spectra, critical localization of eigenstates, and arrange in so-called
quasibands \cite{Prunele01,Prunele02,Bandres16,Vignolo16,Macia17}. The coexistence of localized and delocalized
eigenstates in aperiodic systems can involve a so-called mobility edge which marks the transition energy
separating the different classes of states \cite{Mott87} or can become manifest in an intermediate phase of
interdispersed localized and delocalized states without
mobility edge. The paradigm for exploring mixed localization-delocalization behaviour in one spatial dimension is the
Aubry-Andr\'{e} quasiperiodic model \cite{Aubry80}, whose modifications and generalizations have been extensively
explored \cite{Sarma88,Sarma90,Hashimoto92,Biddle09,Biddle10,Biddle11,Ganeshan15,Liu15,Li15,Li17,Monthus19,
Li20,Roy21,He22,Goncalves22,Goncalves23,Vu23} in particular in recent years.
It should be noted that (infinitely extended) quasiperiodic or even disordered lattice
models are homogeneous in the sense that their overall structure remains the same throughout
the lattice, i.e. it does not depend on the region of the lattice.


\noindent
Quasicrystals do not possess global symmetries but a plethora of local symmetries \cite{Morfonios14,Roentgen19}. 
The impact of the presence of local symmetries in general settings, i.e. beyond the paradigm of quasicrystals,
has been explored recently for both continuous and discrete one-dimensional systems 
\cite{Kalozoumis14a,Kalozoumis13a,Kalozoumis13b,Morfonios17,Schmelcher17,Zambetakis16}. Local symmetries
allow to classify resonances in wave scattering \cite{Kalozoumis13a,Kalozoumis13b} and
enhance the transfer efficiency in lattices \cite{Morfonios20}. Signatures of local symmetries
have been observed experimentally in both lossy acoustic waveguides \cite{Kalozoumis15} and coupled
photonic wave guide lattices \cite{Schmitt20}. A typical spectral feature in the presence of local
symmetries is the localization of eigenstates on the corresponding
local symmetry specific domains of a given lattice (see in particular \cite{Roentgen19}).
The underlying mechanism of this steered localization behaviour has been identified \cite{Schmelcher24}
as the isospectrality of the isolated symmetry-related subdomains i.e. applying a
reflection or translation operation to a given lattice domain does not alter its eigenvalues. As a consequence
we obtain pairwise degenerate eigenvalues that split linearly with an increasing coupling strength of these
symmetry-related subdomains. This is the key ingredient to the present work: we elevate these properties to a working
principle that generates a new category of lattices being composed of coupled isospectral cells, beyond
the notion of local symmetries. The isospectral cells are parametrized by phase angles, or shortly, phases:
for $K \times K$-cells there is $\frac{K(K-1)}{2}$ such phases. For isospectrally patterned lattices (IPL)
these phases vary from cell to cell in a controlled manner across the lattice. While this opens the doorway
of many possible such variations and resulting lattice setups we focus here, as a first exploration
of an IPL, on cells characterized by a single phase. Our lattice covers a fixed phase interval by 'moving' across
the lattice. By definition, these specific IPL are finite and of inhomogeneous character.
The aim of this work is to perform a first spectral analysis of these IPL on the basis of their eigenvalues and eigenstates.
Each of the resulting energy bands is divided into three distinct branches. Two of those branches
show single center (SC) localized states based on a characteristic localization length which results from the competition
of the coupling and the (discrete) phase gradient relating different isospectral cells.
They are separated by a finite system localization delocalization crossover (FLDC) from a third branch
for which the delocalized eigenstates extend over the complete lattice.
We determine the behaviour of the fraction of (de-)localized states with varying phase gradient
and coupling strength as well as number of lattice sites.

\paragraph{Setup and Hamiltonian -} 
According to \cite{Schmelcher24} it is the isospectrality of symmetry-related (isolated) subdomains and the
resulting pairwise degeneracy of eigenvalues which underlie the observed localization of eigenstates
on locally symmetric domains. We therefore consider lattices that consist of isospectral
cells ${\mathbf{A}}_{m}, m \in \{1,...,N\}$ coupled via off-diagonal blocks ${\mathbf{C}}_{m}$, which leads to 
the following Hamiltonian
\vspace*{-0.5cm}

\begin{eqnarray}
{\cal{H}} &=& \sum_{m=1}^{N} \left(\ket{m} \bra{m} \otimes \mathbf{A}_{m} \right)
\label{eq1} \\ \nonumber
& + & \sum_{m=1}^{N-1} \left(\ket{m+1} \bra{m} \otimes \mathbf{C}_{m} + h.c. \right) 
\end{eqnarray}

\noindent
reminescent of the dividing of the state space into internal and external degrees of freedom.
Here $N$ is the number of cells and $N_s$ will be used in the following for the
overall number of lattice sites.
An immediate way of ensuring that the cells $\mathbf{A}_{m}$ are isospectral is to choose them
as orthogonal (or in general unitary) transformations of a diagonal matrix ${\mathbf{D}}$, i.e. we have
${\mathbf{A}}_{m} = {\mathbf{O}}_{\phi_m} {\mathbf{D}} {\mathbf{O}}_{\phi_m}^{-1}$, where
$\phi_m$ indicates the (set of) angles specifying the transformation. In general 
${\mathbf{A}}_{m},{\mathbf{C}}_m$ are $K \times K$ finite sublattices (submatrices)
residing on the diagonal and
couplings of the lattice. For the purpose of providing evidence of the richness of the
spectral properties of ${\cal{H}}$, we specialize to the case $K=2$, resulting in a
single angle $\phi$, and to ${\mathbf{C}}={\mathbf{C}}_m = \frac{\epsilon}{2} \left(\sigma_x + i \sigma_y \right)$ using
open boundary conditions for our lattice. The inhomogeneity of our lattice is implemented by our choice
of the values $\phi_m$: we choose an equidistant grid of angles centered around the value
$\frac{\pi}{4}$. This choice is motivated by the fact that the eigenvectors of
${\mathbf{A}}_{\frac{\pi}{4}}$ are, independent of $\mathbf{D}$, maximally delocalized in the $m-$th cell, providing a
distinct starting-point for the control of localization versus delocalization on the lattice consisting
of many coupled cells. The complete angular (or phase) range covered by the lattice reads then
$[\frac{\pi}{4}-\frac{L}{2},\frac{\pi}{4}+\frac{L}{2}]$ with $L = \frac{\pi}{4} \cdot \frac{1}{L_f}$
where $L_f$ is a scaling factor of the phase range of the lattice, and we have
$\phi_m = \frac{\pi}{4} - \frac{L}{2} + \frac{m-1}{N-1} L, m \in \{1,...,N\}$. Our lattice possesses, by 
construction, an inversion symmetry around its center $\phi = \frac{\pi}{4}$. We note, that for the limiting
case $L_f \rightarrow \infty$ we obtain a (finite) periodic lattice with the unit-cell being ${\mathbf{A}}_{\frac{\pi}{4}}$
whereas for $L_f = 1.0$ the lattice covers the angular range $[\frac{\pi}{8},\frac{3\pi}{8}]$.
We use the notion of localized states as being (non-fragmented) states residing within our (finite) IPL
and not reaching the boundaries of the lattice whereas delocalized states extend over the complete lattice.
Equally we refer to the crossover point in energy where localized states turn into delocalized states
as a FLDC edge. Our main focus will be on the weak to intermediate coupling regime $0 < \epsilon < 0.5$.

\begin{figure}[H]
\centering
\includegraphics[width=8cm,height=7cm]{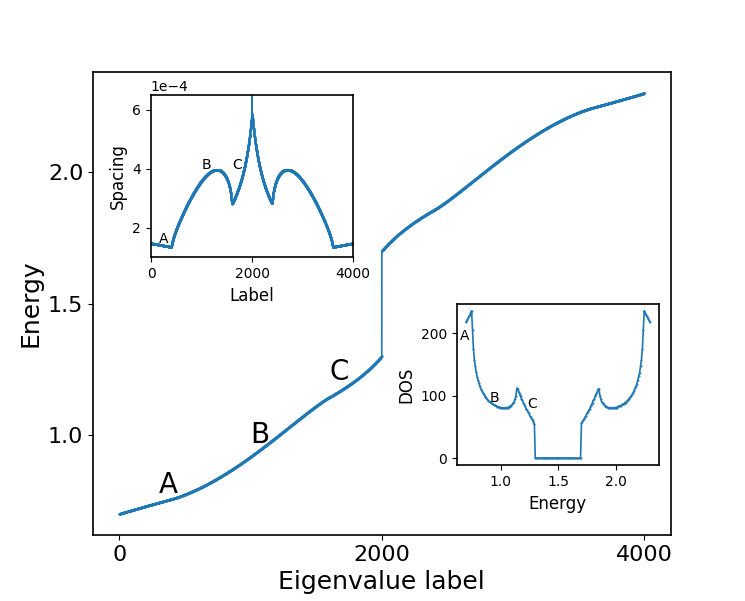}
\caption{Main figure: energy eigenvalue spectrum of the equidistant $\phi$ lattice for $d_1=1,d_2=2, L_f=1.0, \epsilon=0.3,
N_s=4002$, where $d_1,d_2$ are the diagonal values of ${\mathbf{D}}$. Upper left inset: the corresponding energy level spacing.
Lower right inset: the density of states for $N_s=16002$. The labels A,B,C mark the three distinct energetical regimes 
of the band which reflect itself correspondingly in the level spacing and the density of states.} 
\label{Fig:1}
\end{figure}

\paragraph{Phenomenology of the eigenvalue spectrum -} We first analyze the eigenvalue spectrum belonging to the Hamiltonian
${\cal{H}}$ in eq.(\ref{eq1}) for a IPL with several thousand sites and $L_f=1.0$ for a coupling 
strength $\epsilon = 0.3$ and diagonal values $d_1=1,d_2=2$ of $\mathbf{D}$, as shown in Fig.\ref{Fig:1}. 
We observe two bands separated by a band gap \cite{footnote1}, our subsequent
statements holding essentially for both bands. Each band can be divided into three distinct energy domains marked
as A, B, C in Fig.\ref{Fig:1}, which correspond to the lower, middle and upper energy domain
of the lower band. Obviously, the energy eigenvalues with increasing degree of excitation
show a prominent difference from the cosine-dispersion relation of the (monomer) periodic tight-binding case. 
In particular we witness close to the edges of the bands an approximately linear behaviour of the energies.
To work this out in more detail, the upper left inset in Fig.\ref{Fig:1} shows the spectrum of the eigenvalue spacing
clearly exposing three domains with qualitatively different behaviour for each band. While the region A
shows a linearly decreasing spacing, region B exhibits a highly nonlinear and nonmonotonic dependence,
whereas region C displays a very peaked close to linear behaviour. The three domains can also be identified
in the density of states shown as the lower right inset in Fig. \ref{Fig:1}. In region A we observe
a high density of states which is even increasing within this domain and followed by a steep decline
of the density of states in region B with partial recovery for higher energies, and finally, in region
C, we observe an approximately linear decrease. This behaviour persists qualitatively with varying coupling
strength, noting that the energy gap between the two bands closes for $\epsilon = 0.5$. With increasing value
of $L_f$ the covered phase interval shrinks and consequently the sizes of the domains A and B also shrink.

\begin{figure}[H]
\centering
\includegraphics[width=8cm,height=7cm]{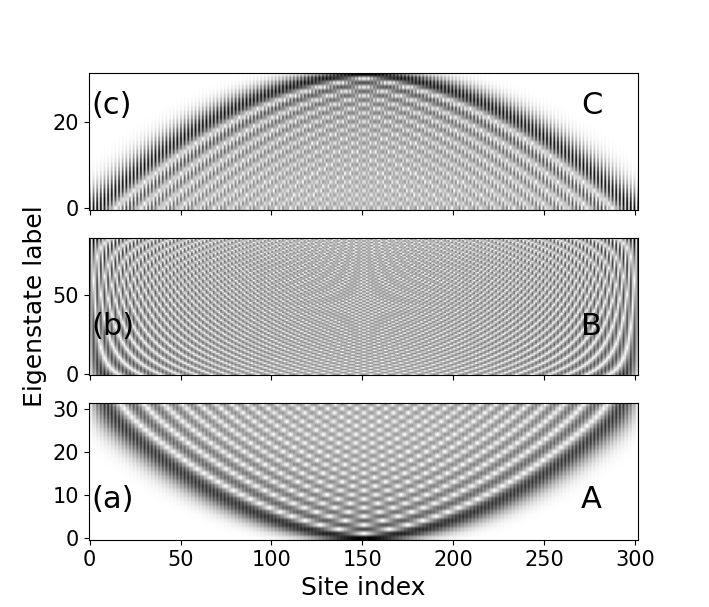}
\caption{Grey scale eigenstate map showing the absolute values of all eigenstate components for the lower band of the 
spectrum of the lattice with $d_1=1,d_2=2, L_f=1.0, \epsilon=0.3, N_s=302$. Note that the grey scale has been
renormalized for each eigenstate row. The sequence of eigenstates is divided into three
domains (a,b,c) according to the energy domains A,B,C of the lower band in Fig.\ref{Fig:1}. Note that the counting of
the eigenstate labels is reset to zero within each domain providing a total of $151$ eigenstate profiles.}
\label{Fig:2}
\end{figure}

\paragraph{Eigenstate analysis: localized versus delocalized states -} 
We have identified above three distinct domains within each energy band which we analyze now in terms
of the behaviour of their eigenstates. Fig.\ref{Fig:2} shows a greyscale eigenstate map, i.e. the magnitude of
the eigenstate components, for the complete first band. The subfigures \ref{Fig:2} (a,b,c) correspond to the
domains A,B,C in the first energy band in Fig.\ref{Fig:1}, respectively. We observe that in the domain A (Fig.\ref{Fig:2}(a))
all eigenstates are localized around a single center
and increasingly spread with increasing degree of excitation. In domain B the
eigenstates are delocalized over the complete lattice, and, finally, in domain C localization takes over again
and the eigenstates become increasingly SC localized with increasing degree of excitation. This behaviour persist
for varying coupling strength, in particular also for weak couplings, and for varying
angular interval $L$ (see below for quantitative statements) and is therefore
of generic character. It repeats in the second upper band. We therefore encounter within each band two
FLDC edges marking the transition from localized to delocalized eigenstates.

\begin{figure}[H]
\centering
\includegraphics[width=8cm,height=7cm]{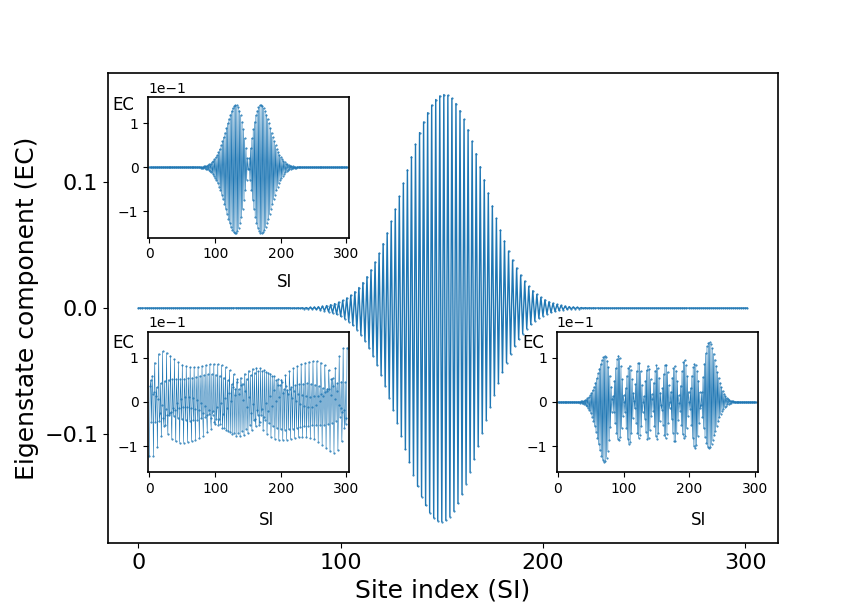}
\caption{Localized states: the ground (main figure), first excited (top left inset) as well as tenth excited
(bottom right inset) state of the first band in the domain A (see Figs.\ref{Fig:1},\ref{Fig:2}).
A delocalized state is shown for comparison (lower left inset) that belongs to domain B.
Parameters are the same as in Fig.\ref{Fig:2}.}
\label{Fig:3}
\end{figure}

Let us inspect the eigenstate profiles in some more detail with the aim to understand the origin of our observed
SC localization.
Fig. \ref{Fig:3} shows, for the same parameter values as in Fig. \ref{Fig:2}, the ground state as well as the first
and tenth excited states in the first band, thereby observing the increasing spreading of the eigenstates.

While our lattice consists of $N_s=302$ sites, the ground state shows a localization length of the order of $50$ sites,
whose origin and mechanism we shall analyze in the following. A closer inspection reveals that the envelope
of the ground state is very well described by a Gaussian wave function. The fast oscillations from
site to site can be attributed to the fact that ${\mathbf{A}}_{\frac{\pi}{4}}$ possesses 
the eigenvectors $(1,-1),(1,1)$.
Resultingly, a variational ansatz for the ground state wave function reads as follows

\begin{equation}
\ket{\Psi} = {\cal{N}} \sum_n \text{exp} \left(- \alpha \left(n-n_0 \right)^2 \right) \ket{n} \otimes (1,-1)
\label{eq2}
\end{equation}

\noindent
where ${\cal{N}} = \frac{1}{\sqrt{2}} 
\left( \sum_n {\text{exp}} \left(- 2 \alpha \left(n-n_0 \right)^2 \right) \right)^{-\frac{1}{2}}$
is the normalization constant and $\alpha$ is a variational parameter to be determined by minimizing
the corresponding energy $E = \bra{\Psi} {\cal{H}} \ket{\Psi}$. Evaluating this expectation value 
involves approximating the summations by continuous integrals and leads to the final result

\begin{eqnarray}
E &=& \frac{1}{2} \left( d_1 + d_2 \right) + \frac{1}{2} \left( d_1-d_2 \right) {\text{exp}} \left( - \frac{1}{4 \beta} \right)
\label{eq3}
\\ \nonumber
&& - \hspace*{0.1cm} \epsilon \hspace*{0.1cm} {\text{exp}} \left( - \frac{\alpha}{2} \right)
\end{eqnarray}

\noindent
where $\beta = \frac{8 \alpha N^2}{\pi^2}$. Note that $\frac{\pi}{4 N L_f}$ is the phase gradient across our lattice 
with $L_f = 1.0$ in the present case. The two competing second and third terms in the energy eq.(\ref{eq3}) 
are due to the phase change across the cells on the diagonal and the off-diagonal coupling terms, respectively.

\noindent
Varying $\alpha$ there exists a single minimum which amounts,
for our specific case, to $\alpha_0 \approx 6.7 \cdot 10^{-3}$.
The resulting energy agrees with the corresponding numerical value within one tenth of a per mill.
The full width half maximum for these analytical considerations is $41$ sites as compared to the
numerical value of approximately $44$ sites. The observed localization behaviour therefore emerges in our lattice
due to the competition in energy between the phase gradient among the isospectral cells
and the coupling between the cells. 

\noindent
The eigenstates at the upper band edge can be obtained in a similar manner. From the above analysis, the position
of the FLDC edge occurs at $n_{mob} \approx C \cdot \frac{N^2}{\sigma^2}$ where $\sigma^2$ is the
variance of the Gaussian ground state and $C$ being a constant of order one. 

\begin{figure}[H]
\centering
\includegraphics[width=8cm,height=6cm]{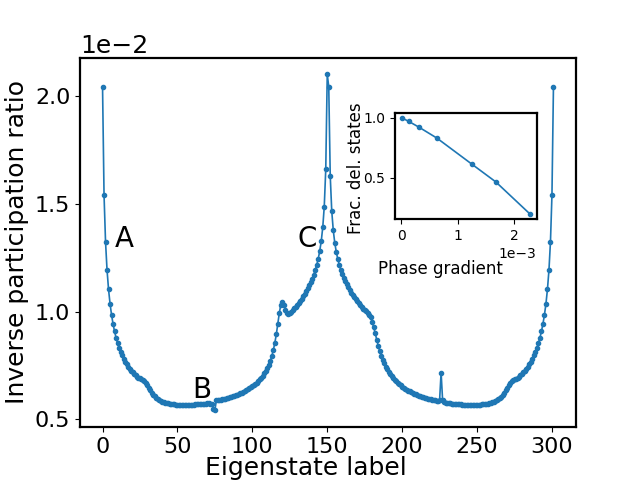}
\caption{Main figure: The inverse participation ratio for all eigenstates across both 
bands. Parameters are the same as in Fig.\ref{Fig:2}. Inset: The fraction of delocalized states
with varying phase gradient for a fixed lattice size $N_s$. Labels A,B,C correspond to the different
energy domains (see Fig. \ref{Fig:1}).}
\label{Fig:4}
\end{figure}

\noindent
To quantify the degree of localization we determine the
inverse participation ratio (IPR) for the complete spectrum of eigenstates,
which is defined as $r = \sum_{i=1}^{N} |\psi_i|^4 \in [N^{-1},1]$. The maximal value for the IPR is one for an 
eigenvector localized on a single site of the chain and the minimal value $\frac{1}{N}$ is encountered for 
a state which is uniformly extended over the chain. As expected, the IPR is large in the domains A and C of 
localized states whereas it is smallest in the regime B of delocalized states where a plateau of low values
is encountered. According to the increasing delocalization in regime A with increasing degree of excitation
the IPR peaks strongly for the ground state and low excitations but then rapidly decays when approaching
the regime B. The reverse happens at the upper edge of the first band. A central moment analysis allows equally
to distinguish between the different domains A,B and C with all odd moments being zero. 
It is instructive to compare the SC localization with the localization occuring for a 
random lattice for the same coupling strength $\epsilon = 0.3$ but random onsite binary disorder with the
values corresponding to the eigenvalues of our IPL $2 \times 2$-sublattices, namely $d_1=1, d_2=2$. The IPR values of the random
multi-centered eigenstates are typically significantly larger than $0.1$ which has to be compared with the much smaller values
of less than $0.02$ for our IPL, see Fig. \ref{Fig:4}. This indicates the much larger localization
length scale of the IPL SC localized eigenstates compared to the random lattice multi-centered eigenstates.

\noindent
Some discussion concerning the fraction of SC localized vs. delocalized states is in order.
For a sufficiently large value of $L_f$ practically all eigenstates are delocalized. With
a decreasing value of $L_f$ for a fixed lattice size, i.e. an increasing phase gradient, 
we observe an approximately linear 
decrease of the fraction of delocalized states, see the inset of Fig.\ref{Fig:4}.
For e.g. $L_f=1.0$ corresponding to a phase gradient of $1.2 \cdot 10^{-3}$ 
(for the above other parameter values) 40 $\%$ of the eigenstates become SC localized.
The fraction of (de-)localized states is independent of the coupling strength $\epsilon$.
While our simulations naturally address finite lattices there is a systematic way
of increasing the number of lattice sites $N_s$. Within this procedure we keep the coupling
constant $\epsilon$ and in particular the covered total
angular interval $L$ fixed and step by step increase the number of lattice points $N_s$.
It should be noted that this course of action maintains the inhomogeneous character of
our lattice and might be reminescent of a continuum limit which should be carefully
distinguished from a systematic increase of the lattice size in the case of homogeneous systems.  
We hereby observe that the fraction of (de-)localized states is independent of the lattice size
which we have verified by varying $N_s$ over three orders of magnitude. For delocalized eigenstates
close to the band center in regime $B$ we obtain for the IPR $r \propto N_s^{-1}$, as expected, where as
for localized states close to the band edges in regimes $A,C$ we observe $r \propto N_s^{-0.5}$.
Averaging over the eigenstates within the (localized) regimes $A,C$ yields a scaling behaviour 
according to $r_m \propto N_s^{-0.84}$ whereas in the (delocalized) regime $B$ one obtains
the well-known $r_m \propto N_s^{-1}$ scaling, where $r_m$ is the corresponding mean IPR. Our 
crossover from localized to delocalized states is robust against disorder, both for the coupling
and for the eigenvalues in the isospectral cells, up to the several percent
level, from which on localized structural changes in the eigenstates are manifest.

\noindent
\paragraph{Conclusions and Perspectives -} 
IPLs open a new pathway of systematically exploring and controlling a designed mixture and crossover
between SC localized and delocalized states as well as devising the corresponding 
finite system localization delocalization crossover edges. The underlying mechanism
of the (de-)localization is the competition between the underlying phase gradient across
different isospectral cells of the lattice and the coupling among those cells.
While we have been focusing here on an inhomogeneous IPL setup which consists of 
a grid of equidistantly separated phases covering a phase interval of the order of a 
single period, i.e. of the order of $2 \pi$, there is a plethora of different possibilities
to design other isospectrally patterned lattices. For example, fixing the phase gradient
and increasing the number of lattice sites such that the phase interval covers several periods
or even extending it to infinity, establishes interesting IPL with potential novel properties
depending, e.g., on the rationality of the phase gradient. Another future line of investigation
would be to study the topological properties of IPL with varying coupling strength between
the cells. A non-constant phase gradient across the lattice would add to this richness of setups.
Going beyond rotations in two dimensions i.e. shaping the isospectral cells in the higher dimensional
angular space intriguing phase structures will represent an interesting and promising avenue to be pursued.

\noindent
Experimental platforms that might be suited to realize IPLs could be integrated photonic
waveguide lattices \cite{Szameit10,Kremer21} or optical lattice/tweezer-based ultracold atomic
systems which offer an astounding control of both external as well internal atomic degrees of
freedom \cite{Bloch08,Browaeys20}. Specifically the recently established synthetic lattices of
laser-coupled atomic momentum modes \cite{An21} could be promising candidates for the realization of IPL
where injection spectroscopy \cite{Paladugu24} is available for probing energy spectra.

\noindent
\paragraph{Acknowledgments -} 
This work has been supported by the Cluster of Excellence “Advanced Imaging of Matter” of the Deutsche
Forschungsgemeinschaft (DFG)-EXC 2056, Project ID No. 390715994. P.S. acknowledges the unique
atmosphere and scientific discussions at ITAMP, Cambridge, in the framework of an extended scientific visit.
P.S. thanks M. R\"ontgen for helpful discussions and Th. Posske for a careful reading of the 
manuscript. Many fruitful interactions and discussions with F.K. Diakonos are very much appreciated.


\begin{thebibliography}{99}
\bibitem{Hamermesh89} M. Hamermesh, Group Theory and Its Applications to Physical Problems,
Dover Books on Physics and Chemistry, 1989. 
\bibitem{Ashcroft76} N.W. Ashcroft and N.D. Mermin, Solid State Physics, Holt-Saunders, 1976.
\bibitem{Singleton01} J. Singleton, Band Theory and Electronic Properties of Solids, Oxford Master Series
in Condensed Matter Physics, Oxford University Press 2001.
\bibitem{Lee85} P.A. Lee and T. V. Ramakrishnan, Disordered elctronic systems, Rev.Mod.Phys. 57, 287 (1985).
\bibitem{Brandes03} T. Brandes, and S. Kettemann, The Anderson Transition and its Ramifications - 
Localisation, Quantum Interference, and Interactions, Lect.Not.Phys. 630, Berlin: Springer Verlag (2003).
\bibitem{Macia09} E. Maci\'{a} Barber, Aperiodic Structures in Condensed Matter, Fundamentals and Applications,
Series in Condensed Matter Physics, CRC Press 2009.
\bibitem{Macia21} E. Maci\'{a}-Barber, Quasicrystals, Fundamentals and Applications, CRC Press 2021.
\bibitem{Shechtman84} D. Shechtman, I. Blech, D. Gratias, and J. W. Cahn, Metallic phase with long-range orientational
order and no translational symmetry, Phys.Rev.Lett. 53, 1951 (1984).
\bibitem{Suck02} J.B. Suck, M. Schreiber, P. H\"aussler, Quasicrystals: An Introduction to Structure,
Physical Properties and Applications, Springer Science \& Business Media 2002.
\bibitem{Janssen86} T. Janssen, Crystallography of quasi-crystals, Act. Crystall. A 42, 261 (1986).
\bibitem{Berger93} C. Berger, T. Grenet, P. Lindqvist, P. Lanco, J. Grieco, G. Fourcaudot and F. Cyrot-Lackmann,
The new AlPdRe icosahedral phase: Towards universal electronic behaviour for quasicrystals?, Solid State Commun. 87, 977 (1993).
\bibitem{Vieira05} A.P. Vieira, Low-Energy Properties of Aperiodic Quantum Spin Chains, Phys.Rev.Lett. 94, 077201 (2005).
\bibitem{Tanese14} D. Tanese, E. Gurevich, F. Baboux, T. Jacqmin, A. Lemaitre, E. Galopin, I. Sagnes, A. Amo,
J. Bloch and E. Akkermans, Fractal Energy Spectrum of a Polariton Gas in a Fibonacci Quasiperiodic Potential,
Phys.Rev.Lett. 112, 146404 (2014).
\bibitem{Jagannathan21} A. Jagannathan, The Fibonacci quasicrystal: Case study of hidden dimensions and multifractality,
Rev.Mod.Phys. 93, 045001 (2021).
\bibitem{Morfonios14} C. Morfonios, P. Schmelcher, P.A. Kalozoumis and F.K. Diakonos, Local symmetry dynamics in 
one-dimensional aperiodic lattices: a numerical study, Nonl.Dyn. 78, 71 (2014).
\bibitem{Prunele01} E. de Prunel\'{e} and X. Bouju, Fibonacchi, Koch and Penrose structures: Spectrum of finite 
subsystems in three-dimensional space, Phys.Stat.Sol.(b) 225, 95 (2001).
\bibitem{Prunele02} E. de Prunel\'{e}, Penrose structures: Gap labeling and geometry, Phys.Rev.B 66, 094202 (2002).
\bibitem{Bandres16} M.A. Bandres, M.C. Rechtsman and M. Segev, Topological photonic quasicrystals: Fractal
topological spectrum and protected transport, Phys.Rev.X 6, 011016 (2016).
\bibitem{Vignolo16} P. Vignolo, M. Bellec, J. B\"ohm, A. Camara, J.M. Gambaudo, U. Kuhl and F. Mortessagne,
Energy landscape in a Penrose tiling, Phys.Rev.B 93, 075141 (2016).
\bibitem{Macia17} E. Maci\'{a}, Clustering resonance effects in the electronic energy spectrum of tridiagonal
Fibonacchi quasicrystals, Phys.Stat.Sol.B 254, 1700078 (2017).
\bibitem{Mott87} N. Mott, The mobility edge since 1967, J.Phys. C: Solid State Physics, 20, 3075 (1987).
\bibitem{Aubry80} A. Aubry and G. Andr\'{e}, Analyticity breaking and Anderson localization in incommensurate lattices,
Ann.Isr.Phys.Soc. 3 (133), 18 (1980).
\bibitem{Sarma88} S. Das Sarma, S. He, and X. C. Xie, Mobility edge in a model
one-dimensional potential, Phys.Rev.Lett. 61, 2144 (1988).
\bibitem{Sarma90} S. Das Sarma, S. He, and X. C. Xie, Localization, mobility edges, and metal-insulator transition in a class of
one-dimensional slowly varying deterministic potentials, Phys.Rev. B 41, 5544 (1990).
\bibitem{Hashimoto92} Y. Hashimoto, K. Niizeki, and Y. Okabe, A finite-size scaling analysis of the 
localization properties of one-dimensional quasiperiodic systems, J.Phys.A 25, 5211 (1992).
\bibitem{Biddle09} J. Biddle, B. Wang, D. J. Priour, and S. Das Sarma, Localization
in one-dimensional incommensurate lattices beyond the Aubry-Andr\'{e} model, Phys.Rev.A 80, 021603(R) (2009).
\bibitem{Biddle10} J. Biddle and S. Das Sarma, Predicted mobility edges in 
one-dimensional incommensurate optical lattices: An exactly solvable model of Anderson localization, 
Phys.Rev.Lett. 104, 070601 (2010).
\bibitem{Biddle11} J. Biddle, D. J. Priour, B. Wang, and S. Das Sarma, Localization
in one-dimensional lattices with non-nearest-neighbor hopping: Generalized Anderson and Aubry-Andr\'{e} models, Phys.Rev.B
83, 075105 (2011).
\bibitem{Ganeshan15} S. Ganeshan, J. H. Pixley, and S. Das Sarma, Nearest neighbor
tight binding models with an exact mobility edge in one dimension, Phys.Rev.Lett. 114, 146601 (2015).
\bibitem{Liu15} F. Liu, S. Gosh and Y.D. Chong, Localization and adiabatic pumping in a generalized Aubry-Andr\'{e}-Harper model,
Phys.Rev.B 91, 014108 (2015).
\bibitem{Li15} X. Li, S. Ganeshan, J.H. Pixley and S. Das Sarma, Many-Body Localization and Quantum 
Nonergodicity in a Model with a Single-Particle Mobility Edge, Phys.Rev.Lett. 115, 186601 (2015).
\bibitem{Li17} X. Li, X.P. Li, and S. Das Sarma, Mobility edges in one-dimensional bichromatic incommensurate potentials, 
Phys.Rev.B 96, 085119 (2017).
\bibitem{Monthus19} C. Monthus, Multifractality in the generalized Aubry-Andr\'{e} quasiperiodic localization
model with power-law hoppings or power-law Fourier coefficients, Fractals 27, 1950007 (2019).
\bibitem{Li20} X. Li and S. Das Sarma, Mobility edge and intermediate phase in one-dimensional incommensurate lattice potentials,
Phys.Rev.B 101, 064203 (2020).
\bibitem{Roy21} S. Roy, T. Mishra, B. Tanatar, and S. Basu, Reentrant localization transition in a quasiperiodic chain, 
Phys. Rev. Lett. 126, 106803 (2021).
\bibitem{He22} Y. He, S. Xia, D.G. Angelakis, D. Song, Z. Chen and D. Leykam,
Persistent homology analysis of a generalized Aubry-Andr\'{e}-Harper model, Phys.Rev.B 106, 054210 (2022).
\bibitem{Goncalves22} M. Gon\c{c}alves, B. Amorim, E. V. Castro, and P. Ribeiro, Hidden 
dualities in 1D quasiperiodic lattice models, SciPost Phys. 13, 046 (2022).
\bibitem{Goncalves23} M. Gon\c{c}alves, B. Amorim, E. Castro and P. Ribeiro,
Renormalization group theory of one-dimensional quasiperiodic lattice models with commensurate approximants,
Phys.Rev.B 108, L100201 (2023).
\bibitem{Vu23} D. D. Vu and S. Das Sarma, Generic mobility edges in several
classes of duality-breaking one-dimensional quasiperiodic potentials, Phys.Rev.B 107, 224206 (2023).
\bibitem{Roentgen19} M. R\"ontgen, C.V. Morfonios, R. Wang, L. Dal Negro and P. Schmelcher,
Local symmetry theory of resonator structures for the real-space control of edge states in binary aperiodic chains,
Phys.Rev. B 99, 214201 (2019).
\bibitem{Kalozoumis14a} P.A. Kalozoumis, C. Morfonios, F.K. Diakonos and P. Schmelcher,
Invariant of broken discrete symmetries, Phys.Rev.Lett. 113, 050403 (2014).
\bibitem{Kalozoumis13a} P.A. Kalozoumis, C. Morfonios, F.K. Diakonos and P. Schmelcher,
Local symmetries in one-dimensional quantum scattering, Phys.Rev.A 87, 032113 (2013)
\bibitem{Kalozoumis13b} P.A. Kalozoumis, C. Morfonios, N. Palaiodimopoulos, F.K. Diakonos and P. Schmelcher,
Local symmetries and perfect transmission in aperiodic photonic multilayers, Phys.Rev.A 88, 033857 (2013).
\bibitem{Morfonios17} C. Morfonios, P.A. Kalozoumis, F.K. Diakonos and P. Schmelcher,
Nonlocal discrete continuity and invariant currents in locally symmetric effective
Schr\"odinger arrays, Ann.Phys. 385, 623 (2017).
\bibitem{Schmelcher17} P. Schmelcher, S. Kr\"onke and F.K. Diakonos,
Dynamics of local symmetry correlators for interacting many-particle systems, J.Chem.Phys. 146, 044116 (2017).
\bibitem{Zambetakis16} V.E. Zambetakis, M.K. Diakonou, P.A. Kalozoumis, F.K. Diakonos, C.V. Morfonios and P. Schmelcher,
Invariant current approach to wave propagation in locally symmetric structures, J.Phys.A 49, 195304 (2016).
\bibitem{Morfonios20} C.V. Morfonios, M. R\"ontgen, F.K. Diakonos and P. Schmelcher,
Transfer efficiency enhancement and eigenstate properties in locally symmetric 
disordered finite chains, Ann.Phys. 418, 168163 (2020).
\bibitem{Kalozoumis15} P.A. Kalozoumis, O. Richoux, F. K. Diakonos, G. Theocharis, and P. Schmelcher,
Invariant currents in lossy acoustic waveguides with complete local symmetry, Phys.Rev.B 92, 014303 (2015).
\bibitem{Schmitt20} N. Schmitt, S. Weimann, C.V. Morfonios, M. R\"ontgen, M. Heinrich, P. Schmelcher and A. Szameit, 
Observation of Local Symmetry in Photonic Systems, Las.Phot.Rev. 14, 1900222 (2020).
\bibitem{Schmelcher24} P. Schmelcher, Degenerate subspace localization and local symmetries, Phys.Rev.Res. 6, 023188 (2024).
\bibitem{footnote1} Note, that we are using the terminology of a band, for reasons of convenience, although our
setup is indeed non-periodic.
\bibitem{Szameit10} A. Szameit and S. Nolte, Discrete optics in femtosecond-laser-written photonic structures,
J.Phys. B 43, 163001 (2010).
\bibitem{Kremer21} M. Kremer, L.J. Maczewsky, M. Heinrich and A. Szameit,
Topological effects in integrated photonic waveguide structures,
Opt.Mat.Expr. 11, 1014 (2021).
\bibitem{Bloch08} I. Bloch, J. Dalibard, and W. Zwerger,
Many-body physics with ultracold gases, Rev.Mod.Phys. 80, 885 (2008).
\bibitem{Browaeys20} A. Browaeys and T. Lahaye, Many-body physics with individually controlled Rydberg atoms,
Nat.Phys. 16, 132 (2020).
\bibitem{An21} F.A. An, K. Padavi\u{c}, E.J. Meier, S. Hegde, S. Ganeshan, J.H. Pixley, S. Vishveshwara and B. Gadway,
Interactions and Mobility Edges: Observing the Generalized Aubry-Andr\'{e} Model, Phys.Rev.Lett. 126, 040603 (2021).
\bibitem{Paladugu24} S.N.M. Paladugu, T. Chen, F.A. An, B. Yan and B. Gadway, Commun. Phys. 7, 39 (2024).
\end{thebibliography}
\end{document}